\documentclass[12pt,preprint]{aastex}
\newcommand{\bea}{\begin{eqnarray} }
\newcommand{\eea}{\end{eqnarray}}

\begin{document}

\title{Obscuring Material around Seyfert Nuclei with Starbursts}

\author{Keiichi WADA$^1$}
\affil{National Astronomical Observatory of Japan, Mitaka, Tokyo
181-8588, Japan\\
E-mail: wada.keiichi@nao.ac.jp}
\altaffiltext{1}{CASA, University of Colorado, 389UCB, Boulder, CO 80309-389}

\author{Colin A. NORMAN$^2$}
\affil{Johns Hopkins University, Baltimore, MD 21218\\
E-mail: norman@stsci.edu}
\altaffiltext{2}{Space Telescope Science Institute, Baltimore, MD 21218}


\begin{abstract}
The structure of obscuring matter in the environment of active
galactic nuclei with associated nuclear starbursts is investigated
using 3-D hydrodynamical simulations. Simple analytical
estimates suggest that the obscuring matter with energy feedback from
supernovae has a torus-like structure with a radius of several tens of
parsecs and a scale height of $\sim 10$ pc. These estimates are
confirmed by the fully non-linear numerical simulations, in which the
multi-phase inhomogeneous interstellar matter and its interaction with
the supernovae are consistently followed.  The globally stable,
torus-like structure is highly inhomogeneous and turbulent. To achieve
the high column densities ($ \gtrsim 10^{24}$ cm$^{-2}$) as suggested by
observations of some Seyfert 2 galaxies with nuclear starbursts, the
viewing angle should be larger than about 70$^\circ$ from the pole-on
for a $10^8 M_\odot$ massive black hole. Due to the inhomogeneous
internal structure of the torus, the observed column density is
sensitive to the line-of-sight, and it fluctuates by a factor of order
$\sim 100$. The covering fraction for $N > 10^{23}$ cm$^{-2}$ is 
about 0.4.  The average accretion rate toward $R < 1$ pc is $\sim 0.4
M_\odot$ yr$^{-1}$, which is boosted to twice that in the model without the
energy feedback.
\end{abstract}




%


\section{INTRODUCTION}

Optically thick obscuring molecular tori have been postulated to
explain various properties of active galactic nuclei (AGN), especially
the two major categories of the AGN, namely
type 1 and type 2.  However, the true structure and the formation and
maintenance mechanism of the torus have not yet been understood either
theoretically or observationally.

Recently a number of observations have suggested a new aspect of
Seyfert 2 galaxies (Sy2).  It is pointed that there are possibly two
types of Sy2s: one is a classic Sy2, an obscured Seyfert 1
nucleus. The other type II is a Sy2 with a nuclear starburst
\citep{cid95,gra97,gon01,lev01}.  In the latter case it is believed
that a nuclear starburst is associated with the obscuring material
whose scale is $R < 100 $ pc.  These observations imply that the
classic picture of the unified model for AGNs, in which the diversity
of the Seyfert galaxies is explained only by the geometrical effect of
a geometrically and optically thick torus, may not be applicable to
some fraction of the Sy2s (see also \citet{maio95}).  More generally
we propose here that there is an additional parameter in the unified
models, namely, the strength of the nuclear starburst. It is this
nuclear starburst and the mass of the black hole (BH) that determines the
geometry of the obscuring torus.

While discussing the X-ray background, Fabian et al. (1998) proposed
that low-luminosity AGN could be obscured by nuclear starbursts within
the inner $\sim 100$ pc.  They suggested that supernovae (SNe) from a nuclear
starburst can provide the energy to boost the scale height of
circumnuclear clouds and so obscure the nucleus.
For simplicity, they assumed that the material sits in an isothermal
sphere, and the gravitational effects of the central BH are
ignored.
This idea seems to be consistent with the recent observations
of Sy2s with nuclear starbursts, 
Here we investigate the
model in more detail.  The problem is actually very complicated. We
need to solve the 3-D inhomogeneous structure and
dynamics of the ISM in the combined gravitational potential of the
central massive BH, the central stellar system and the central
massive gas distribution.  We must handle the radiative cooling not
only for the hot gas, but also for the cold gas,
 and include the feedback interaction between SNe
and the inhomogeneous ISM.  Numerical simulations are powerful tools
for this kind of complicated problem.

Recently we presented high-resolution, 3-D
 hydrodynamical modeling of the ISM in the central
 hundred pc region in galaxies, taking into account self-gravity of
 the gas, radiative cooling, and energy feedback
 from SNe \citep{wad01b,wada01} (hereafter WN01 and W01).
 In this {\it Letter}, we apply this numerical
 method to the gas dynamics around the central massive BH, and
 study the starburst-AGN connection in Seyfert galaxies.

%
\section{GEOMETRY OF OBSCURING MATERIAL AROUND THE CENTRAL ENGINE}
%
In this section, we discuss the possible structure of the obscuring
material around the central massive BH under the influence of
the SN explosions detonating in the central massive gas cloud.
The obscuring material will be inhomogeneous, and multi-phase and the
geometrical thickness is assumed to be supported by its internal
turbulent motion caused by energy input from SNe.

We assume the turbulent energy dissipation in a disk around the
central massive BH is in equilibrium with the energy input
from SNe.  In a unit time and volume, the energy
balance is
 \bea
 \frac{\rho_g v_t^2}{\tau_d} = \eta s_\star E_s = \eta \alpha \frac{v_c}{r} \rho_g E_s,
\label{eqn: 1}
 \eea
 where $v_t$ is turbulent velocity of the gas, $\tau_d$ is the
 dissipational time scale of the turbulence, $E_s$ is the total energy
 injected by a SN, $\eta$ is a heating efficiency per unit
 mass that represents how much energy from SNe is converted to
 kinetic energy of the ISM, and $s_\star$ is star formation rate per
 unit volume and time.  We can assume the star formation rate is
 proportional to the dynamical time scale in the disk namely
$  s_\star \sim \alpha \Omega_c \Sigma_g h^{-1} = \alpha {v_c}/{r} \rho_g,$ 
where $v_c$ is the circular velocity, $r$ is a
radius, $h$ is the scale height of the disk, $\Sigma_g$ is surface
density, and $\alpha$ is a constant ($\alpha \sim 0.02$ is suggested
for a global star formation in spiral galaxies rate by Kennicutt
1998).
The dissipational time scale $\tau_d$ can be simply assumed as
$ \tau_d = \xi h/v_t, $ 
where $\xi$ is a constant of order unity. Therefore
eq.(\ref{eqn: 1}) is
\bea
 \frac{v_t^3}{\xi h} = \eta \alpha \frac{v_c}{r} E_s. \label{eqn:
5}
 \eea
Assuming  hydrostatic equilibrium for the vertical direction of the disk,
i.e. $\rho_g v_t^2 \sim \rho_g g h$, we have
$
 v_t^2 \sim  {G  M_{\rm BH}}{r^{-3}} h^2,  \label{eqn: 6} 
$
for the case that the central massive BH dominates the gravitational potential (regime I),
or
$
 v_t^2 \sim  \pi G \Sigma_\star h,    \label{eqn: 7} 
$
if the stellar disk or bulge potential dominates the potential. Here
$\Sigma_\star$ is surface mass density of the stellar system (regime II).
For the regime I, $ {h}/{r}  \sim {v_t}/{v_c}$ 
and for the regime II, $ {h}/{r}  \sim {v_t^2}/{v_c^2}$.
Therefore from eq.(\ref{eqn: 5}), we have
$
 h \sim (\xi \eta \alpha E_s)^{1/2} r v_c^{-1},  \label{eqn: 10}
$
for the regime I, or for the regime II,
$
 h \sim (\xi \eta \alpha E_s)^2 r v_c^{-4}.  \label{eqn: 11}
$
Using $v_{c,I} = (G M_{\rm BH}/ r)^{1/2}$ for the regime I ($r < r_0$), or  $v_{c,II} = (\pi G \Sigma_\star)^{1/2} r^{1/2}$ for 
the regime II ($r>r_0$), the disk scale height
in each regime is
\bea
 h_I(r) &=& (\xi \eta \alpha E_s )^{1/2} G^{-1/2} M_{\rm BH}^{-1/2} r^{3/2},   \; r < r_0   \label{eqn: 14a}
\eea
and
\bea
 h_{II}(r) &=& (\xi \eta \alpha E_s )^2 (\pi G \Sigma_\star)^{-2} r^{-1},  \;\;\;\;\;\;\;\;\; r \geq r_0  \label{eqn: 14b}
\eea
where $r_0$ is determined from $v_{c,I}(r_0) = v_{c,II}(r_0)$,  and it is
$
 r_0  =  (M_{\rm BH}/\pi \Sigma_\star )^{1/2} 
\sim 56 \, {\rm pc} \;\; M_8^{1/2} \Sigma_{\star,4}^{-1/2},  \label{eqn: 16}
$
where the BH mass, $M_8 \equiv 10^8 M_\odot$ and 
 $\Sigma_{\star,4} = \Sigma_{\star}/10^4 M_\odot {\rm pc}^{-2}$.
Therefore the maximum scale height at each regime is
 \bea
h_{0,I}(r_0) &=& (\xi \eta \alpha E_s )^{1/2} G^{-1/2} M_{\rm BH}^{1/4} 
(\pi \Sigma_\star)^{-3/4} \nonumber \\
&\sim&  22 \, {\rm pc} \;\;
(\xi_1 \alpha_2 \eta_{-3} E_{51})^{1/2} M_8^{1/4} (\pi \Sigma_{\star,4})^{-3/4} \, , \label{eqn: 17} \\
h_{0,II}(r_0) &=& (\xi \eta \alpha E_s )^2 G^{-2} M_{\rm BH}^{-1/2} 
(\pi \Sigma_{\star})^{-3/2} \nonumber \\
&\sim&  1\, {\rm pc} \; (\xi_1 \alpha_2 \eta_{-3} E_{51})^2 
M_8^{-1/2} \Sigma_{\star,4}^{-3/2}  \, ,
\eea
where  $\xi_1 \equiv \xi/1.0$, $\alpha_2 \equiv \alpha/0.02$,
$\eta_{-3} = 10^{-3} M_\odot^{-1}$, the total gas mass $M_{g,8} = M_g/10^8 M_\odot$. 
The scale height can be also expressed as
$h_{0,I}(r_0) \sim 35 \, {\rm pc} \;\; SFR_1^{1/2} r_{6}^{3/2} (\xi_1 E_{51}/M_{g,8})^{1/2},$ with the star formation rate
$SFR_1 \equiv 1 M_\odot$ yr$^{-1}$ and $r_{6} = r/60$ pc.
Here we assume the Salpeter's IMF.
Note that $h_{0,II} > h_{0,I}$, if $\alpha_2 \gtrsim 0.08$ for
$\xi_1, \eta_{-3},M_8, \Sigma_{\star,4},$ and $E_{51}$.
Eqns (\ref{eqn: 14a}) and (\ref{eqn: 14b}) show that the disk has a torus-like structure (it looks more like a {\it concave lens} rather than a {\it bagel}),
 and its opening angle $\Phi$ is 
$ \tan \Phi \sim r_0/h_{0,I} 
\propto (\xi \eta \alpha E_s)^{-1/2} M_{\rm BH}^{1/4} \,,
$
for $h_{0,II} < h_{0,I}$.
Therefore $\Phi \sim  70^\circ$ for $\xi_1, \eta_{-3}, \alpha_2,
M_8$, and $\Sigma_{\star,4}$.
For the case that nuclear starburst is very active,
 e.g.  $SFR_1 =$ 10 and 100, we have smaller opening angles,
i.e. $\Phi \sim 50^\circ$ and $20^\circ$, respectively.

If the star formation originates {\it via} gravitational instability
in the gaseous disk, stars are not formed near the massive black
hole. Assuming the Toomre's stability criterion $\Sigma_{crit} =
\Omega_c c_s/\pi G = c_s G^{-1/2} M_{\rm BH}^{1/2} r^{-3/2}/\pi$,
where $c_s$ is the sound velocity, the critical radius $r_c$ for which
the disk with a uniform surface density is unstable is $ r_c \sim
\left[ {c_s^2 r_0^4 M_{\rm BH}}/({M_g^2 G}) \right]^{1/3}, $ then $ r_c
\sim 3 \, {\rm pc} \; c_1^{2/3} r_{6}^{4/3} M_8^{1/3} M_{g,8}^{-2/3},
\label{eqn: 19} $ where $c_1 = c_s/(1$ km s$^{-1})$.  The disk scale
height for the stable disk would be very thin inside $r = r_c$,
otherwise its shape would be represented by eqns (\ref{eqn: 14a}) and
(\ref{eqn: 14b}).  Therefore the gas around the central massive black
hole with the nuclear starburst naturally forms a geometrically thick
obscuring structure around the AGN, depending on the mass of the black
hole, the gas mass, and the strength of the burst.

For the case of $h_{0,I} \gg h_{0,II}$, the average column density toward the nucleus as a function of the viewing angle $\phi$ ($\phi = 0$ is pole-on),
 $\tilde{N}(\phi) \sim \langle \rho_g \rangle L(\phi)$, where $L(\phi)$ is 
the path length in the torus, is approximately 
$
\tilde{N}(\phi) \sim 5 \times 10^{24} {\rm cm}^{-2} \; 
 M_{g,8} r_{6}^{-1} h_{22}^{-1} \Gamma(\phi),
$
where $h_{22}\equiv h/22$ pc. $\Gamma(\phi)$ is a function that determines the path length in the
torus,
$
 \Gamma(\phi) \sim (1-2 \tan^{-1}\phi)^{1/2}/\sin \phi ,
$
for $ r_0/h_{0,I} \lesssim \tan \phi \lesssim r_0/h_{0,II}$, or
$
 \Gamma(\phi) \sim \cos^{-1} \phi ,
$
for $ \tan \phi > r_0/h_{0,II}$.
$\Gamma \sim 0.8$ for $ \phi \sim 80^\circ$. If the star formation rate
is ten times larger and $h_{0,I} < h_{0,II}$, 
$\tilde{N}(\phi)$ is smaller by a factor of two.

%
\section{NUMERICAL MODELING OF THE OBSCURING GAS IN THE GALACTIC CENTER}
%
The argument in \S 2 is
basically an order-of-magnitude estimate, and, consequently, detailed
structure and evolution of the ISM in the galactic center should be
verified by a fully non-linear, time-dependent numerical simulations.
The star-forming massive gas should be an
inhomogeneous with a velocity field that is turbulent (see WN01 and W01).
SN explosions in such inhomogeneous
system cannot be simply parameterized.
The evolution of the 
SNe remnants will not be spherically symmetric, and should not
be expressed by a simple analytic solution.  As a result, it is not
straightforward to estimate how much energy of a SN explosion
is converted to the kinetic energy of the inhomogeneous, non-stational
ISM.  The energy dissipation rate as a function of time and position
in the non-stationary turbulent gas disk needs to be explicitly
calculated.
In this section, we use a state-of-the-art numerical technique to
clarify the evolution and structure of the gas in the star-forming
region around the central massive BH.

\subsection{Numerical Method and Model}

The numerical methods are the same as those described in WN01 and W01.
Here we briefly summarize them.
\setcounter{footnote}{0} We solve the same equations as eqns.(1)--(4) 
in W01 numerically
in 3-D to simulate the evolution of a rotating ISM in 
a fixed gravitational potential. Here the external potential force in 
the momentum conservation equation (eqn. (2) in W01) is
$\nabla \Phi_{\rm ext} + \nabla \Phi_{\rm BH}$, 
where 
the time-independent
external potential is $\Phi_{\rm ext} \equiv -(27/4)^{1/2}v_c^2/(r^2+
a^2)^{1/2}$ with the core radius $a = 10$ pc and
the maximum rotational velocity $v_c = 100$ km s$^{-1}$. 
The central BH potential is
$\Phi_{\rm BH} \equiv -GM_{\rm BH}/(r^2 + b^2)^{1/2}$ with $b=1$ pc.
We also assume a cooling function $\Lambda(T_g) $ $(20 < T_g < 10^8
{\rm K})$ with Solar metallicity and a heating due to
photoelectric heating, $\Gamma_{\rm UV}$ and due to energy feedback
from SNe, $\Gamma_{\rm SN}$.  We assume a uniform UV radiation field,
which is ten times larger than the local UV field.
The hydrodynamic part of the basic equations is solved by AUSM
 (Advection Upstream Splitting Method)\citep{LS}. 
We use $256^2 \times 128$ Cartesian grid points covering a $64^2
\times 32$ pc$^3$ region around the galactic center (the
spatial resolution is 0.25 pc).  The Poission equation is solved 
using the FFT and the convolution method.
The initial condition is an axisymmetric and rotationally
supported thin disk with a uniform density profile (thickness is 2.5
pc) and a total gas mass of $M_g = 5\times 10^7 M_\odot$
Random density and temperature fluctuations, which are 
less than 1 \% of the unperturbed values, 
are added to the initial disk.

SN explosions are assumed to occur at random positions
 in the region of 
$|x|,|y| < 51$ pc and $|z| < 4$ pc. The average SN
rate is $\sim$ 1 yr$^{-1}$, which is corresponds to $\alpha \sim 0.3$ at $r = 20$ pc.
The energy of $10^{51}$ ergs is instantaneously
injected into a single cell as thermal energy. 
The 3D evolution of blast waves caused by SNe in an inhomogeneous and non-stationary medium with global
rotation is followed explicitly, taking into account the radiative cooling.
Therefore the evolution of the SNRs, e.g. the duration and structure 
of the SNe, depends on the gas density distribution around the SN.

%
\subsection{Numerical Results}
%

Figure 1 (a), (b) and (c) show volume-rendering representations of the
density distribution of the gas around the massive BH in a
quasi-stable state ($t=1.6$ Myr) from three different viewing angles.
In Fig. 1(a), the nuclear region can be seen, whereas the nucleus is
obscured in Fig. 1(b) and 1(c).
The right half of the Fig 1 (b) is a
cross-section of the thick disk, and the inner structure of the thick
disk is clear; the scale height of the disk is smaller in the central
region than in the outer region as expected in \S 2, and the density
structure is very inhomogeneous.  The three panels in Fig. 1 together
show that the gases around the central massive BH form a
torus-like structure.  The inhomogeneous, filamentary structure is
quasi-stable.  That is, the local inhomogeneous structure is
time-dependent, but the global torus-like morphology does not
significantly change during several rotational period.

In Fig. 2, we plot line-of-sight column density of the gas as a
function of the viewing angle (90$^\circ$ is edge-on).  The column
density is higher in edge-on view.  To achieve $N > 10^{23-24}$
cm$^{-2}$, which is suggested in observations of Sy2s with nuclear
starbursts \citep{lev01}, the viewing angle should be larger
than $\sim 70^\circ$ in this model.  It should be noted, however, that
the column density has large (about two orders of magnitude)
fluctuations for the same viewing angle. This is caused by the
inhomogeneous internal structure of the torus, which shows a
Log-Normal distribution function (see W01).  A factor of 100 change in
the average column density corresponds to about $\pm 20^\circ$ in the
viewing angle. We find that a fraction of the solid angle, for which
the nucleus is obscured, is 0.21, 0.37, 0.46, and 0.55 for $N > 10^{24},
10^{23},$ $10^{22}$, and $10^{21}$ cm$^{-2}$, respectively. 
These numbers are a
factor 2--3 smaller to explain the ratio of Sy2s to Sy1s, which is
about 4 \citep{maio95b}, based on the `strict' unified model.

From eqns. (\ref{eqn: 17}), 
we can expect the scale height of the disk of for $\alpha = 0.3$ is
$h \sim 16$ pc at $r = 20$ pc and $h \sim 6 $ pc at  at $r = 10$ pc. 
This consistent with
 the numerical result. The column density predicted in \S 2
for $\phi \sim 80^\circ$ fit the simulation.
However, Fig.2 shows that the column density depends more strongly 
on the viewing angle than the analytical prediction. This is because the
average density distribution for a $z$-direction in the torus is not constant,
contrary to the assumption in \S 2.

Fig. 3 shows the time evolution of the gas mass inside $r = 1$ pc for
the models with and without energy feedback. The average accretion
rate is ${\dot M}_g \sim 0.4 M_\odot$ yr$^{-1}$, which is roughly
twice that in the model without feedback. The turbulent viscous time
scale in this system is $t_{\rm tub} \sim r^2/\nu_{\rm tub}$ with
viscous coefficient $\nu_{\rm tub}$.  Using $\nu_{\rm tub} \sim v_t
\times l_t$, where $l_t$ is the largest eddy size which is about the
scale height of the disk ($\sim 10$ pc), the turbulent viscous time
scale is $t_{\rm tub} \sim 10^8 (v_t/10 {\rm km} \, {\rm s}^{-1})^{-1}
$ yr.  This is the same order of the numerical result, i.e. $M_g/{\dot
M}_g \sim 5\times 10^7 M_\odot/0.4M_\odot {\rm yr}^{-1}$.

%
\section{DISCUSSION}
%
The estimate in \S 2 suggested that 
the opening angle positively correlates with $M_{\rm BH}$.
As $M_{\rm BH}$ increases, the critical radius for the
gravitational instability is also larger.
This implies that low luminosity (i.e.
small $M_{\rm BH}$) AGNs are obscured for larger fraction of the
viewing angle than the high luminosity AGN, if, quite reasonably, the
mass accretion rate has a positive dependence on the mass of the
central engine.  In other words, luminous type 2 objects, such as type
2 quasars are less frequently observed in the optical than type 1
quasars. They could be observed mainly X-rays, and this is consistent
with observations \citep{norman01,leh01,akiyama00}.

\citet{kro88} (KB) claimed that stirring by stellar processes is never
strong enough to compete with energy dissipation in the clumpy
torus. One should note, however, that in their order of magnitude
estimate, a much smaller, thicker torus ($\sim 1$ pc) is assumed than
in our extended disk model. Therefore KB need large velocity
dispersions ($> 200$ km s$^{-1}$) to keep the disk thick, because 
thick disks with $h/r \sim 1$ require $v_t/v_c \sim 1$. 
The velocity dispersion for $h/r \sim 1$ can be less than 50 km s$^{-1}$
for $r \gtrsim 30$ pc. The smaller velocity dispersion is favorable, because
the energy dissipation rate is smaller (see eq.(2)).
KB also pointed out that a
luminosity-to-gas-mass ratio, $L/M_g$, for the stellar stirring model
would be a factor 100 larger than the observed values.  This is also
not the case for a 100 pc scale obscuring material. For the analytical
model presented here, $L/M_g \sim r^{-3/2}$. In fact, a much larger
molecular gas mass is expected on such a scale (e.g.
\citet{kohno99}). Recent IR and X-ray observations of Seyfert nuclei
(e.g.  \citet{gra97}), on the other hand, suggest a 100 pc scale
extended obscuring material. This picture is consistent with our model
of the extended starburst supported obscuring region in which case we
still obtain a geometrically thick disk with velocity dispersions
$\lesssim 50$ km s$^{-1}$.
Note also
Pier \& Krolik (1992) and Ohsuga \& Umemura (2001) for the effects of
radiation pressure on supporting the obscuring torus.

The present model suggests that the ratio of Sy2s to Sy1s cannot be
explained only by the orientation of the viewer relative to the torus.
In Sy1s, the star formation rate in the nuclear region would be
smaller than in Sy2s.  The strict unified model, in this sense, should
be modified by including the strength of the nuclear starburst.

%
\acknowledgments 
%
We thank our colleagues Roberto Cid-Fernandez, Tim Heckman, Julian
Krolik and Nancy Levenson for useful and stimulating discussions.
Numerical computations were carried out on Fujitsu VPP5000 at NAOJ.  
KW is supported by Grant-in-Aids for Scientific Research 
(no. 12740128) of JSPS.

\newpage

\figcaption{3-D density distribution represented by a volume-rendering
technique. 
The colors represent the relative opacity based on the line-of-sight column
density, but it does not represent the absolute opacity or any
physical value.  The regions colored red or yellow are more optically
thick than the blue regions. 
The right half of (b) is a surface section.}
\figcaption{Column density vs. viewing angle. The column density
toward the nucleus is calculated for the density distribution of 
a model shown in Fig. 1 every $\theta \sim 4^\circ$ and 
$4^\circ$ for azumithal direction.}
\figcaption{Time evolution of the gas mass inside $R < 1$ pc for
two models (with and without energy feedback). The solid line
represents the mass accretion rate $0.3 M_\odot$ yr$^{-1}$.}

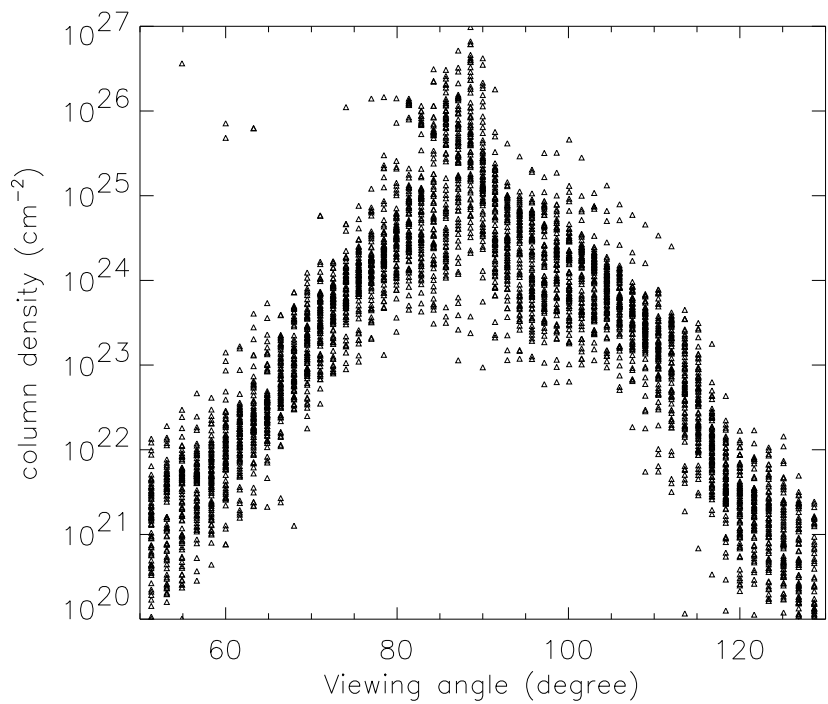
\end{document}